\documentclass{elsart}
\newcommand{\bea}{\begin{eqnarray}}
\newcommand{\eea}{\end{eqnarray}}

\newcommand{\vvec}{\vec{v}}
\usepackage{graphics}
\begin{document}
\begin{frontmatter}
\title{Networks in life: Scaling properties and eigenvalue spectra}
\author{
I. Farkas$^{\, 1}$,
I. Der\'{e}nyi$^{\, 1,2}$, 
H. Jeong$^{\, 3}$,
Z. N\'{e}da$^{\, 3,4}$,
Z. N. Oltvai$^{\, 5}$,
}
\author{
E. Ravasz$^{\, 3}$,
A. Schubert$^{\, 6}$,
A.-L. Barab\'{a}si$^{\, 3}$, 
T. Vicsek$^{\, 1}$
}

\address{$^1$Department of Biological Physics, E\"otv\"os Lor\'and University,
H-1117 Budapest, Hungary}
\address{$^2$Institut Curie, UMR 168, 26 rue d'Ulm, F-75248 Paris 05, France}
\address{$^3$Department of Physics, University of Notre Dame, Notre
Dame, IN 46556}
\address{$^4$Department of Theoretical Physics, Babes-Bolyai University, RO-3400, Cluj, Romania}
\address{$^5$Department of Pathology, Northwestern University, Medical
School, Chicago, IL 60611}
\address{$^6$Bibliometric Service, Library of the Hungarian Academy of
Sciences, H-1245 Budapest, Hungary}

\begin{abstract}

We analyse growing networks ranging from collaboration graphs of
scientists to the network of similarities defined among
the various transcriptional profiles of living cells.
For the explicit demonstration 
of the scale-free nature and hierarchical organization of these graphs,
a deterministic construction is also used.
We demonstrate the use of determining the eigenvalue spectra of 
sparse random graph models for the categorization of small measured networks.

\end{abstract}
\end{frontmatter}

\section{Introduction}

Numerous natural, social and technological systems
develop large complex structures made up from many similar,
but still specific and individual units connected in a stochastic way.
The simplest approach still rich in details to such phenomena uses
{\it random network models}
built upon ideas from random graph theory and statistical physics.

In this paper we give an overview of the major random graph models
and also show an example for a determistic scale-free network.
We analyse and model the {\it social networks} 
derived from scientific co-authorship data in mathematics and in neuroscience.
For a more detailed description of random networks, we 
suggest the usage of {\it spectral properties}.
Using a {\it molecular biological network} we demonstrate
that the spectral characterization is appropriate for
the categorization of measured networks even in the case of small
systems made up of a few hundred nodes.

\section{Network models}
\subsection{Stochastic graphs}
\label{stochgraphs}

The {\it uncorrelated random graph model} 
of Erd\H os and R\'enyi \,\cite{ErdosRenyi}
treats the network as an assembly of {\it identical units}, where the
number of edges grows quadratically with the size of the system, $N$.
However, in realistic cases, the number of edges grows less rapidly,
e.g., linearly with system size.
The {\it small-world graph}\,\cite{watts1,watts2}
can be created 
by connecting each $1$st, $2$nd, ..., $k$th 
neighbor pair of a one-dimensional periodic lattice 
and then randomly rewiring 
a given portion, $p_{\rm r}$, of the edges
of the original regular graph.
The resulting graph shows the {\it small-world property}: 
neighbors of an arbitrary vertex are often connected to each other, 
as well,
but the number of steps (taken on the edges of the graph) 
connecting two arbitrary vertices tends to be low.
In the {\it stochastic scale-free model},
one starts from $m$ isolated vertices,
and at each time step one new vertex is added by connecting it to $m$
previous vertices. 
For a connection, any previous vertex is chosen with
a probability proportional to its degree, $k_i$: 
$\Pi_i = k_i / \sum_{j=1,N} k_j$.
In the infinite size limit, 
the distribution of degrees converges to a power-law, 
i.e., the system has {\it no characteristic length scale}.

\subsection{Deterministic scale-free model}

In the deterministic scale-free model\,\cite{deterministic} the network
is built iteratively, each iteration repeating and reusing the already
existing network. We start from a single node called the {\it root} of
the network. Next, we add two nodes and connect each of them to the
{\it root} of the network. 
In the {\it n}th time step we add two units identical to the
already existing network (containing $3^{n-1}$ nodes each),
and connect each of the $2^n$ bottom points (vertices) of these two units
to the root point.

Due to the deterministic nature of the model,
the degree distribution of hubs can be obtained exactly.
In the limit, the asymptotic behavior of the degree distribution will
depend only on the degrees of 
{\it hubs}\footnote{
The degrees of non-hub points do never exceed 
the index of the iteration, $n$,
therefore, in the $n\to\infty$ limit, 
their degrees will not modify the asymptotic power-law behavior
of the degree distribution.
}, i.e., vertices having at least one further vertex below themselves.
If we exclude the root point, the number of hubs with
$k=2^{n-i+1}-2$ links after the $n$th iteration will be 
$N_k=(2/3)\, 3^{i}$ ($i=1,...,n-1$). 
With the $N_k$ values known,
a standard tool for obtaining the probability density, 
$p(k)$, of the graph's degrees would be 
the cumulative density function,
$\Phi (k)=N^{-1}\,\sum_{k}^{k_{max}} p(k')$ 
($N$ is the total number of hubs).
However, for a clear illustration of the underlying mathematical idea,
here we will give a slightly different derivation
of the probability density of the graph's degrees.

The above formulas of $k$ and $N_k$
will give $N_k\propto k^{-\log 3/\log 2}$
(which can be simplified to $N_{k/2} = 3 N_k$).
When using this relation as a histogram 
of the probability density, the size of histogram bins 
(i.e., the separation between adjacent $k$ values) is
proportional to $k$ itself. Thus,
in the $N\to\infty$ limit we have

\bea
p(k)\propto k^{-1}\,N_k = k^{-(1+\log 3/\log 2)} \, .
\eea

for the degree distribution and 
$\gamma = 1+\log 3/\log 2$ for the
scaling exponent. 
For further recent proposals of deterministic
scale-free models see Refs. 
\,\cite{dorogovtsev01pseudofractal,jung01geometric}.

\section{Mathematics and neuroscience co-authorship networks}

Social networks have been largely studied in social sciences
\,\cite{wasserman,kochen}. A general feature of these studies is that they are
restricted to rather small systems, and view networks as
static graphs, whose nodes are individuals and links represent various 
social interactions.
Recent statistical physics approaches focus instead on large networks,
searching for universalities both in
the topology of the web and in the dynamics governing it's evolution.
These combined theoretical and empirical results have opened
unsuspected directions in many fields ranging from computer science to biology
\,\cite{watts2,wasserman,asb,alb3,alb4,lawrence,huberman,meta,sole1}.

To illustrate the power of these advances, here
we summarize our results for the collaboration network of scientists.
For each research field one can define a co-authorship
network which reflects the professional links between the scientists.
In this network nodes represent individual scientists,
and two scientists are connected if they have ever published together.
In order to gather information on the topology of a scientific
co-authorship web, ideally, one would need a complete dataset of the
published papers starting from the 
beginnings of the considered discipline until today.
However, computer databases cover at most
the past several decades. Thus any study of this kind needs to be limited to 
only a recent segment of the database, imposing unexpected challenges. 

The databases considered by us contain article titles and authors of all
relevant journals in the field of mathematics (M) and neuro-science (NS),
published in the period 1991-98.
In mathematics our database contains 70,975 different authors and
70,901 papers for an interval spanning eight years. In NS the number of
different authors is 209,293 and the number of published papers is 210,750.

It is also important to mention that recently, Newman also applied modern 
network ideas to collaboration networks \,\cite{newman1,newman2}. 
He studied large databases
focussing on several fields of research over a five-year period, finding that 
collaboration networks
possess all general ingredients of small-world networks \,\cite{newman1}.


\begin{figure}
\resizebox{\textwidth}{!}{\includegraphics{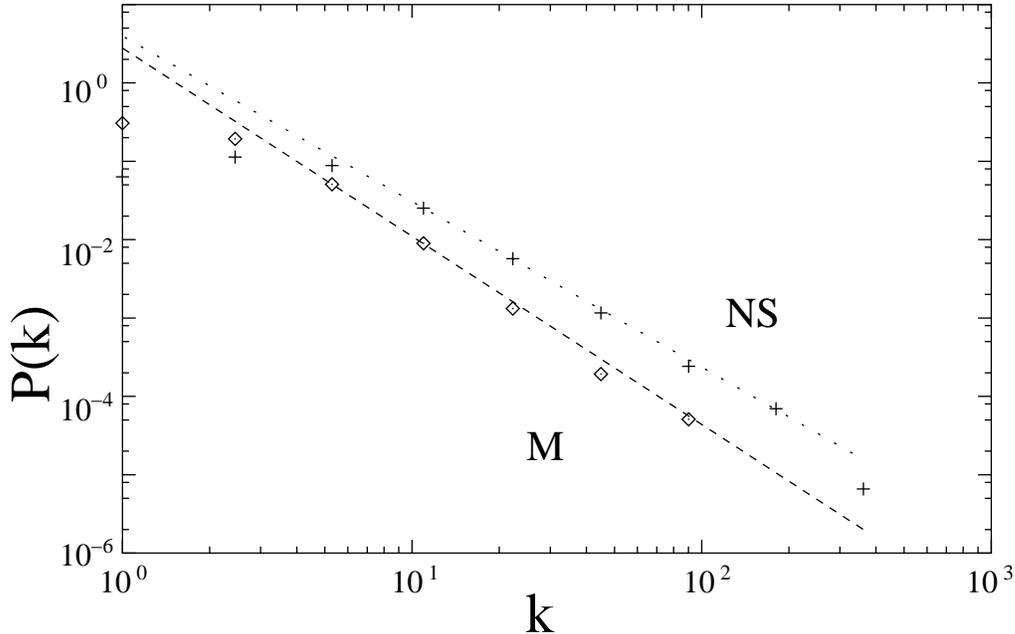}}
\caption{ 
Degree distribution for the M and  NS databases
shown with logarithmic binning
computed from the full dataset, cumulative up to 1998. 
The lines correspond to the best fits
and have the slopes $2.1$ (NS, dotted) and $2.4$ (M, dashed).}
\label{fig:fig2}
\end{figure}

\subsection{Data analysis}
\label{collab_data}

In this section we analyse the topology and dynamics of the
M and NS databases.

A quantity that has been much studied lately for various networks
is the {\it degree distribution}, $P(k)$, giving the probability that a
randomly selected node has $k$
links. The degree
distributions of both the M and NS data indicate that
collaboration networks are scale-free.
The power-law tail is evident from the raw, 
uniformly binned data, but
the scaling regime is better seen on the plot when logarithmic binning
is applied to reduce the noise in the tail (Fig.~\ref{fig:fig2}).

{\it Preferential attachment} is part of all network models aiming to explain 
the emergence of the
inhomogeneous network structure and power law connectivity
distributions \,\cite{dorog1,d3,redner}.
For the networks considered by us preferential attachment appears at
two levels: 

(i) {\em New nodes:} For a new author preferential attachment means that 
it is more likely that his/her first paper will be co-authored with
somebody who 
already has a large number of co-authors (links) than with another
researcher having fewer collaborators.
As a result "old" authors with more links will increase their
number of co-authors at a higher rate than those with fewer links. To
investigate this process in quantitative terms we determined the probability
that an old author with connectivity
$k$ is selected by a new author for co-authorship. This probability
defines the $\Pi(k)$ distribution function. Calling "old authors"
those present up to the last year, and "new authors" those who were added
during  the last year, we computed the change in the number of links,
$\Delta k$, 
for an old author with $k$ links at the beginning of the previous year,
Plotting $\Delta k$ as a function of $k$ gives the function $\Pi(k)$,
describing the nature of preferential attachment involved.
Since measurements are limited to only a finite ($\Delta T=1$ year) interval,
we improve the statistics by plotting the integral of $\Pi(k)$:
\begin{equation}
\kappa(k)=\int_1^k \Pi(k')\, dk'.
\end{equation}

\begin{figure}
\resizebox{\textwidth}{!}{\includegraphics{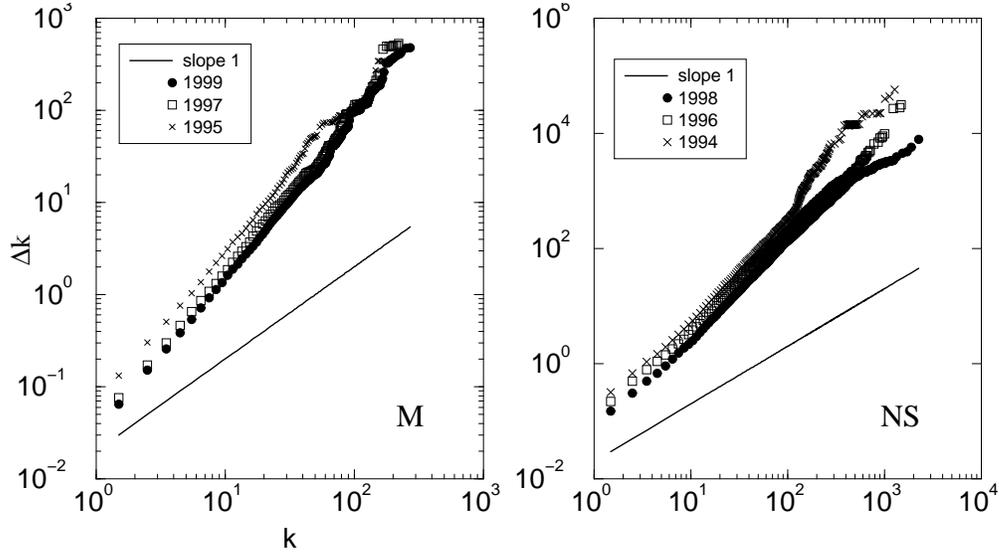}}
\caption{Cumulated preferential attachment 
($\kappa(k)$) of incoming new nodes for the
M and NS databases. In the absence of preferential attachment
$\kappa (k)\sim k$, which shown as a continuous line on the figures.}
\label{fig:fig7}
\end{figure}

If preferential attachment is absent, $\Pi(k)$ should be independent of $k$,
as each node grows independently of its degree, and $\kappa (k)$ is
expected to be linear. As Fig.~\ref{fig:fig7} shows, we find that
$\kappa (k)$ is nonlinear, increasing as $\kappa (k) \sim k^{\nu+1}$,
with the best fits giving $\nu \simeq 0.8$ for M and $\nu \simeq 0.75$ for NS.

\begin{figure}
\resizebox{\textwidth}{!}{\includegraphics{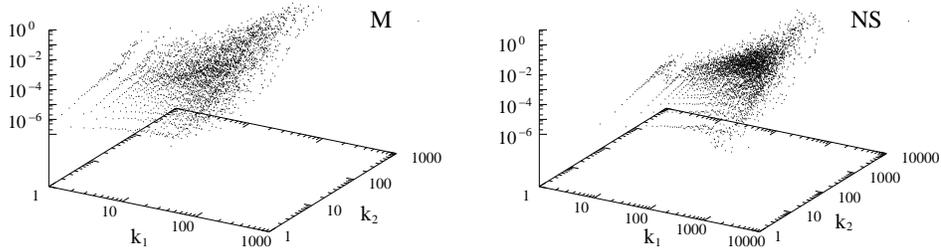}}
\caption{Internal preferential attachment for the M and
NS databases, 3D plots: $\Delta k$ as a function of $k_1$ and $k_2$.
Results computed on the cumulative data in the last considered year.}
\label{fig:fig8}
\end{figure}

(ii) {\em Internal links:} 
As the network evolves, a large number of new links appear
between old nodes representing papers written by
authors already part of the network, but having not collaborated before.
These internal links are also subject to
preferential attachment. We studied the probability $\Pi(k_1,k_2)$
that an old author
with $k_1$ links forms a new link with another old author with
$k_2$ links.
The three dimensional plot of $\Pi(k_1,k_2)$ 
is shown in Fig.\,\ref{fig:fig8}, the overall
behavior indicating that $\Pi(k_1,k_2)$
increases as either $k_1$ or $k_2$ increases.
Analysing in detail our data we also found that 
$\Pi(k_1,k_2)$ for internal links is approximately linear in $k_1 k_2$.

\section{Spectral characterization of random graphs}

Suprisingly, not only human social networks,
one of the highest levels 
of organization among living systems, 
but also food webs\,\cite{foodweb}, 
molecular biological networks\,\cite{meta,wuchty}
and the network of the similarities among various 
genetic programs of a single cell\,\cite{manu}
display small-world and scale-free behavior.

However, until now, most analyses of 
the more realistic models and
the analyses of data sets 
have been confined to the computation of quantities 
which are only loosely connected 
to structural properties:
e.g., degree sequences, shortest connecting 
paths and clustering coefficients. 
Here, we will carry out a more 
detailed analysis
using {\it algebraic tools} intrinsic to random graphs. 
Also, we will show that using algebraic tools
suprisingly {\it small measured networks},
consisting of not more than a few hundred nodes, 
can be successfully {\it classified} into 
one of the realistic network model categories shown above.

\subsection{Definitions and algorithms}

Any graph $G$ can be 
represented by its {\it adjacency matrix}, $A(G)$, 
which is a real symmetric matrix:
$A_{ij}=A_{ji}=1$, if vertices $i$ and $j$ are connected, 
or $0$,	if these two vertices are not connected.

The {\it spectrum of a graph} is the set of eigenvalues of the graph's
adjacency matrix. 
The largest eigenvalue, $\lambda_1$, is also called the
{\it principal eigenvalue} of the graph. 
To illustrate the meaning of the graph's
eigenvalues, consider the
following example. Write each component of a vector $\vvec$ on
the corresponding vertex of the graph: $v_i$ on vertex $i$. 
Next, on every vertex write the sum of the numbers found on
the neighbors of vertex $i$. 
If the resulting vector is a multiple of $\vvec$, then $\vvec$ is an
eigenvector, and the multiplier is the corresponding eigenvalue of the
graph.

\par
The {\it spectral density, $\rho(\lambda)$, of a graph} is the 
density of the eigenvalues of its adjacency matrix.
For a finite system, this can be written as a sum of delta functions

\bea
\rho(\lambda) := {1\over N} \sum_{j=1}^N 
\delta(\lambda -\lambda_j) \, ,
\label{eq_specdens}
\eea

\noindent
which converges to a continuous function with $N\to\infty$.
The spectral density of a graph can be
{\it directly related to the graph's topological features:}
the $k$th moment, $M_k$, of $\rho(\lambda)$ can be written as

\bea
M_k =
{1\over N} \sum_{j=1}^N (\lambda_j)^k =
{1\over N} {\rm Tr} \big( A^k\big) = \
={1\over N} \sum_{i_1,i_2,\cdots ,i_k}
A_{i_1,i_2} A_{i_2,i_3} \cdots A_{i_k,i_1} \, .
\label{eq_mk}
\eea

\noindent
From the topological point of view,
$D_k=NM_k$ is the {\it number of directed paths} (loops) 
of the underlying -- undirected -- graph, that return
to their starting vertex after $k$ steps 
(see Ref. \,\cite{graphspec} for a detailed explanation).

\subsection{Spectral densities of random graph models}

In the infinite system size limit,
the spectral density of an
uncorrelated random graph -- if rescaled as 
$\lambda'=\lambda\, [Np(1-p)]^{-1/2} \propto \lambda\, N^{-1/2}$
-- converges to a semi-circle:

\bea
\rho(\lambda') = \
\cases {
(2\pi)^{-1} \sqrt{4-\lambda'^{\,2}},& if \
$|\lambda' | < 2 \sigma$;\cr 0,&otherwise.} 
\label{eq_semicircle}
\eea

Suprisingly, the ideal semi-circular spectral density is not valid for
any of the above mentioned 
realistic graph models\,\cite{graphspec,golinelli,kahng}).
In the {\it sparse uncorrelated random graph} 
the largest eigenvalue remains constant: 
$\lambda_1/pN\to c$, 
and $\rho(\lambda)$ will be symmetric in the infinite system size
limit. Since the number of isolated clusters of any size will grow
linearly with system size,
the spectral density will contain an infinite number of 
singularities
\footnote{
In a graph with $N$ vertices,
the contribution of one isolated vertex to the spectral density is 
$N^{-1}\delta(\lambda)$, 
and the contribution of an isolated cluster with two vertices is 
$N^{-1}[\delta(\lambda+1)+\delta(\lambda-1)]$.
An isolated cluster with three vertices and two edges will give 
$N^{-1}[\delta(\lambda+\sqrt{2})+\delta(\lambda)+\delta(\lambda-\sqrt{2})]$,
and one with three vertices and three edges adds
$N^{-1}[2\delta(\lambda+1)+\delta(\lambda-2)]$ to the spectral
density.
}
in the limit.
Therefore, in the limit all odd moments ($M_{2k+1}$) converge to
$0$. In other words: the number of all loops with odd length
($D_{2k+1}$) disappear. Since on a tree a path returning to its
starting point must contain any edge an even number of times,
the absence of loops with odd length indicates 
that the sparse uncorrelated random graph becomes more and more
tree-like with $N\to\infty$.
In other words, {\it except for a few shortcuts
a sparse uncorrelated random graph looks like a tree}.

We conclude, that 
-- from the spectrum's point of view --
the {\it high number of triangles
is one of the most basic properties of the small-world model},
and it is preserved much longer, than regularity or periodicity,
if the level of randomness, $p_{\rm r}$, is increased.
Note that the high number of triangles is equivalent to a high
average clustering, $C$, of the graph. 

\unitlength10mm 
\begin{figure}
\resizebox{\textwidth}{!}{\includegraphics{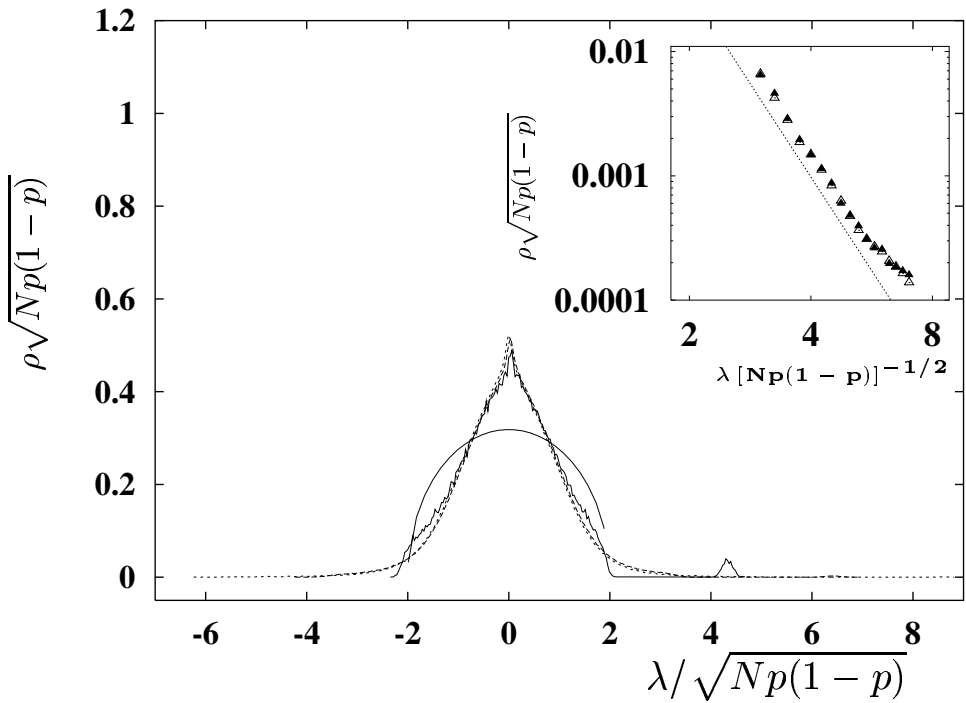}}
\caption[]{
{\bf Main panel:} The average spectral densities of 
scale-free graphs with $m=m_0=5$, and 
$N=100$ (---), $N=1000$ \hbox{(-- --)} 
and $N=7000$ \hbox{(- - -)} vertices.
(For $N=100$ and $N=1000$, the complete spectrum of $1000$ graphs,
and for $N=7000$, the complete spectrum of $25$ graphs was used.)
Another continuous line shows the semi-circular distribution 
for comparison. 
Observe that 
(i) the central part of the scale-free graph's 
spectral density is triangle-like, not semi-circular and 
(ii) the edges show a power-law decay, whereas the
the semi-circular distribution's 
edges decay exponentially\,\cite{Mehta_book}.
{\bf Inset:} The upper and lower tails of $\rho(\lambda)$ 
(open and full triangles) 
for scale-free graphs with $N=40,000$ vertices 
the average degree of a
vertex being $\langle k_i\rangle = 2m = 10$, as before.
Note that both axes are logarithmic and 
$\rho(\lambda)$ has a power-law tail with the same decay rate at both
ends of the spectrum.
The line with the slope $-5$ in this figure is a guide to the eye.
\label{fig:BA}
}
\end{figure}

{\it For $m=m_0=1$, the scale-free graph is a tree} by definition
and its spectrum is symmetric\,\cite{CvetDooSa}.
In the $m>1$ case
$\rho(\lambda)$ consists of several
well distinguishable parts (see Fig.\,\ref{fig:BA}). 
The ``bulk'' part of the spectral density
-- the set of the eigenvalues \{$\lambda_2, ...,\lambda_N$\} --
converges to a symmetric continuous function, which has a
triangle-like shape for $|\lambda'|<1.5$ 
and has power-law tails.

The {\it central part} of the spectral density
{\it converges to a triangle-like shape}
with its top lying well above the semi-circle.
Since the scale-free graph is fully connected by definition,
the increased number of eigenvalues with small magnitudes
cannot be accounted to small isolated clusters.
(All $N$ eigenvalues of a finite-sized 
isolated cluster with $N$ vertices
fall between $-\sqrt{N}$ and $\sqrt{N}$.)
As an explanation, we suggest, that the {\it eigenvectors} 
of these eigenvalues are {\it localized on
a small subset of the graph's vertices}.

The inset of Fig.\,\ref{fig:BA} shows the
{\it tail of the bulk part} of the spectral density
for a graph with $N=20,000$ vertices and 
$100,000$ edges (i.e., $pN=10$). 

\unitlength10mm 
\begin{figure}
\resizebox{\textwidth}{!}{\includegraphics{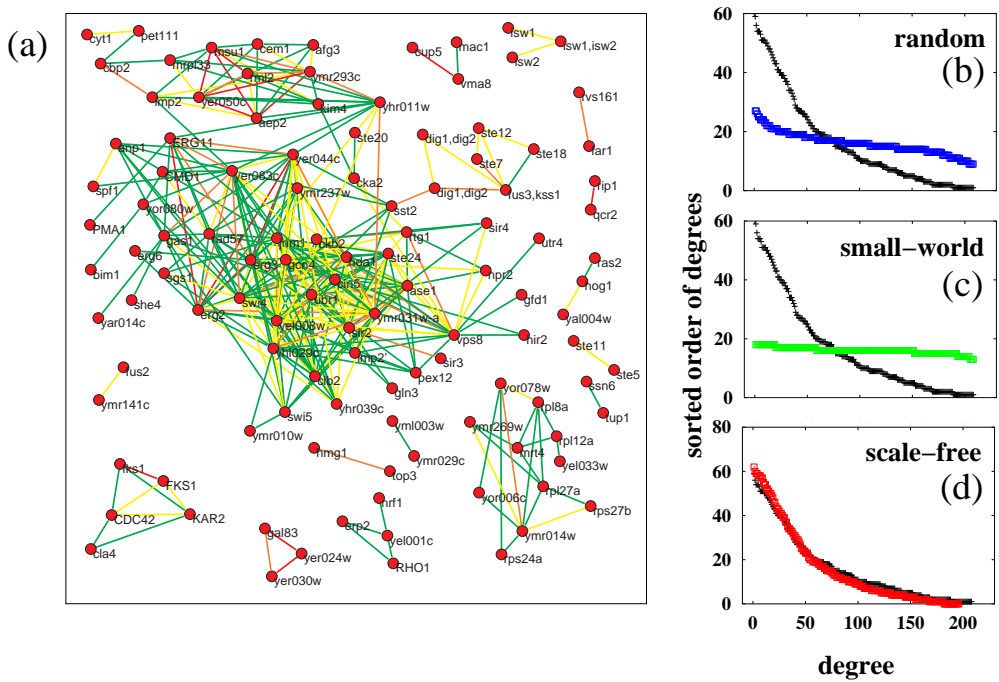}}
\caption[]{
{\bf (a)} The transcriptome 
similarity network of Ref.\,\cite{manu}. Each link is colored
according to its predicted strength: dark links indicate strong
coupling.
{\bf (b-d)} Analysis of the graph's largest component
(the data graph). On each analysis plot
black $+$ signs show data points and open gray
boxes show data points of the three random graph models.
The data graph and all three test graphs 
have the same number of edges and vertices.
The almost complete overlapping of data and test points on subfigure
(d) indicates that the transcriptome similarity graph is scale-free.
\label{fig:transcriptome}
}
\end{figure}

\subsection{Application: classification of a molecular biological
network by spectral methods}

Figure\,\ref{fig:transcriptome} shows a biochemical network derived from
recent gene expression data (see Refs.\,\cite{manu,compendium} 
for an explanation). Nodes of the graph represent individual genetic
programs (also called transcriptomes) of a cell in response 
to various internal or external perturbations. 
A connection between two nodes indicates that according to
the applied similarity search algorithm\,\cite{manu}, 
which compares each pair of
the genetic programs individually, the two indicated transcriptomes
contain a high number of regions with strong correlation ($|C|>0.8$)
of gene expression.

Besides the transcriptome similarity graph, 
Figure \,\ref{fig:transcriptome} also shows its spectral
analysis by comparing the largest component to three idealized test
graphs with the same number of edges and vertices. Note that 
on the inverse participation ratio vs. eigenvalue plots 
the {\it best fit is given by the scale-free graph}, which almost 
completely overlaps with the measured graph's data. The 
principal eigenvalue and the inverse 
participation ratio of the first eigenvector are both high in the
measured and the scale-free graphs, whereas they are both significantly
lower in the two other models. This indicates that the largest
component of the transcriptome similarity graph is scale-free and a
handful of its vertices are structurally dominant.



\begin{thebibliography}{999}

\bibitem{ErdosRenyi} 
P. Erd\H os and A. R\'enyi, 
``On the evolution of random graphs,''
{\it Publ. of the Math. Inst. of the Hung.
Acad. of Sci.} {\bf 5} (1960) 17-61

\bibitem{watts1} D. J. Watts, {\em Small World} (Princeton University Press, Princeton, 1999).

\bibitem{watts2} 
D. J. Watts and S. H. Strogatz, 
``Collective dynamics of `small-world' networks,''
{\it Nature} {\bf 393} (1998) 440-442

\bibitem{deterministic}
A.-L. Barab\'{a}si, E. Ravasz and T. Vicsek,
``Deterministic scale-free networks,''
{\it Physica A}  {\bf 299}, 599-564 (2001)

\bibitem{dorogovtsev01pseudofractal}
S. N. Dorogovtsev, A. V. Goltsev and J. F. F. Mendes, 
(to be published).

\bibitem{jung01geometric}
S. Jung, S. Kim and B. Kahng, 
{\it Phys. Rev. E} {\bf 65} 056101 (2002).

\bibitem{wasserman} S. Wasserman and K. Faust, {\em Social Network Analysis}
		(Cambridge Univ. Press, Cambridge, 1994).

\bibitem{kochen} M. Kochen (ed.) \emph{The Small World} (Ablex, Norwood, NJ, 1989).

\bibitem{asb} L. A. N. Amaral, A. Scala, M. Barth\'el\'emy and H. E. Stanley,
		 Proc. Nat. Acad. Sci. USA {\bf 97}, 11149 (2000).

\bibitem{alb3} R. Albert and A. L. Barab\'asi, Phys. Rev. Lett. {\bf 85}, 5234 (2000).

\bibitem{alb4} R. Albert, H. Jeong and A.L. Barab\'asi, Nature {\bf 400}, 130 (1999).

\bibitem{lawrence} S. Lawrence and C. L. Giles, Nature {\bf 400} 107 (1999).

\bibitem{huberman} B. A. Huberman and L. A. Adamic, Nature {\bf 401} 131 (1999);

\bibitem{meta} H. Jeong et al., Nature {\bf 407}, 
651 (2000).

\bibitem{sole1} R. V. Sol\'e and J. M. Montoya, cond-mat/0011196 (2000);
			
\bibitem{newman1}  M. E. J. Newman, Proc. Nat. Acad. Sci. USA {\bf 98}, 404 (2001).

\bibitem{newman2} M. E. J. Newman {\it Phys.Rev. E} {\bf 64} 016131 (2001).

\bibitem{dorog1} S. N. Dorogovtsev and J. F. F. Mendes, Europhys. Lett. {\bf 52}, 33 (2000).

\bibitem{d3} S. N. Dorogovtsev and J. F. F. Mendes, Phys. Rev. E {\bf 62}, 1842 (2000).

\bibitem{redner}  P. L. Krapivsky, S. Redner and F. Leyvraz, Phys. Rev. Lett {\bf 85}, 4629 (2000);

\bibitem{foodweb}
J. M. Montoya and R. V. Sol\'{e},
``Small World Patterns in Food Webs,''
{\tt cond-mat/0011195}

\bibitem{wuchty}
S. Wuchty, ``Scale-Free Behavior in Protein Domain Networks,''
{\it Mol. Biol. Evol.} {\bf 18} 1694-1702 (2001)

\bibitem{manu}
I. J. Farkas, H. Jeong, T. Vicsek, 
A.-L. Barab\'{a}si and Z. N. Oltvai (to be published)

\bibitem{graphspec} I. J. Farkas, I. Der\'{e}nyi, A.-L. Barab\'{a}si
and T. Vicsek, Phys. Rev. E {\bf 64}, 026704:1-12 (2001)
		 
\bibitem{golinelli}
M. Bauer, O. Golinelli,
``Random incidence matrices: moments of the spectral density,''
{\it J. Stat. Phys.} {\bf 103}, 301-337 (2001)

\bibitem{kahng}
K.-I. Goh, B. Kahng and D. Kim,
``Spectra and eigenvectors of scale-free networks,''
{\it Phys. Rev. E} {\bf 64}, 051903 (2001)

\bibitem{CvetDooSa}
D. M. Cvetkovi\'c, M. Doob and H. Sachs, {\it Spectra of Graphs}
(Berlin, 1980)

\bibitem{Mehta_book}
M. L. Mehta. {\it Random Matrices}. 2nd ed. (Academic, New York, 1991)

\bibitem{compendium}
T. R. Hughes, {\it et.al.}, 
``Functional discovery via a compendium of expression profiles,''
{\it Cell} {\bf 102}, 109-126 (2000)

\end{thebibliography}
\end{document}